\documentclass[aps,pre,twocolumn,english,superscriptaddress,showpacs,floatfix]{revtex4}
\usepackage{amsmath}
\usepackage{psfig,graphics,graphicx}
\usepackage{babel}

\begin{document}

\title{Rapid convergence of time-averaged frequency in phase synchronized
systems}

\author{J\"orn Davidsen}
\email[]{davidsen@mpipks-dresden.mpg.de}
\affiliation{Max-Planck-Institut f\"ur Physik komplexer Systeme,
N\"othnitzer Strasse 38, 01187 Dresden, Germany}
\affiliation{Chemical Physics Theory Group, Department of
Chemistry, University of Toronto, Toronto, ON M5S 3H6, Canada}
\author{Istv\'an Z. Kiss}
\author{John L. Hudson}
\affiliation{Department of Chemical Engineering, 102 Engineers'
Way, University of Virginia, Charlottesville, Virginia 22904-4741}
\author{Raymond Kapral}
\email[]{rkapral@chem.utoronto.ca}
\affiliation{Chemical Physics
Theory Group, Department of Chemistry, University of Toronto,
Toronto, ON M5S 3H6, Canada}

\date{\today}

\begin{abstract}
Numerical and experimental evidence is presented to show that many
phase synchronized systems of non-identical chaotic oscillators,
where the chaotic state is reached through a period-doubling
cascade, show rapid convergence of the time-averaged frequency.
The speed of convergence toward the natural frequency scales as
the inverse of the measurement period. The results also suggest an
explanation for why such chaotic oscillators can be phase
synchronized.
\end{abstract}


\maketitle

\section{Introduction}

The rich collective behavior, including mutual entrainment and
self-synchronization, in systems of coupled oscillators has
been considered by several investigators in the past few years
(see, for example, Refs.~\cite{pikovsky,strogatz00,kuramoto} and
references therein). Recently, a considerable amount of research
has been devoted to the study of coupled \emph{chaotic}
oscillators and, in particular, to the phenomenon of phase
synchronization. Provided that the phase can be defined
\cite{pikovsky97,josic01}, two coupled nonidentical chaotic
oscillators are said to be phase synchronized if their frequencies
are locked but their amplitudes are not
\cite{rosenblum96,pikovsky}. This appears to be a general
phenomenon and it has been observed in systems as diverse as
electrically coupled neurons \cite{elson98,makarenko98},
biomedical systems \cite{schaefer98}, chemical systems
\cite{kiss02}, and spatially extended ecological systems
\cite{blasius99}. Moreover, the potential role of phase
synchronization in brain functions has been explored
\cite{tass03,fitzgerald99}.

The most common theoretical approach to phase synchronization of
chaotic oscillators is based on an analogy with the evolution of
the phase of a periodic oscillator in the presence of external
noise \cite{pikovsky}. This approach leads to the conclusion that
the dynamics of the phase is generally diffusive and the phase
performs a random walk. However, the effective "noise" in such a
description cannot be considered as a Gaussian $\delta$-correlated
noise in all circumstances. In this paper we show that the
effective "noise" exhibits strong temporal (anti-)correlations for
a general class of chaotic attractors. In particular, we present
evidence from simulations and experiments that many phase
synchronized systems of chaotic oscillators where the chaotic
state is reached through a period-doubling cascade show a rapid
convergence of the time-averaged frequency. The speed of
convergence toward the natural frequency scales as the inverse of
the measurement period. This implies that short measurement times
may suffice for reliable determination of frequencies in those
systems.

The outline of the paper is as follows: Section~\ref{sec:back}
presents a sketch of the theoretical background of phase dynamics
in chaotic systems and the expected scaling of the time-averaged
frequency with the observation time. The results of numerical
simulations of globally coupled arrays of R\"ossler oscillators
are presented in Sec.~\ref{sec:sim} where features of the chaotic
attractor leading to rapid convergence of the time-averaged
frequency are identified. Section~\ref{sec:exp} contains a
description and analysis of experimental results on a globally
coupled array of electrochemical oscillators where rapid frequency
convergence is observed.

\section{Theoretical Background} \label{sec:back}
An appropriate definition of the phase for chaotic self-sustained
oscillators can be obtained from the Poincar\'e map of the flow.
Such a map can be constructed if a surface of section in the phase
space of the autonomous continuous-time chaotic system exists
which is crossed transversally by all trajectories of the chaotic
attractor. Then, for each piece of a trajectory between two
crossings of this surface, we define the phase as a linear
function of time:
\begin{equation}
\phi (t) = 2 \pi \frac{t - t_n}{t_{n+1} - t_n} + 2 \pi n,
\end{equation}
for $t_n \leq t < t_{n+1}$. Here, $t_n$ is the time of the $n$th
crossing of the surface of section. The definition is
ambiguous because it depends on the choice of the
Poincar\'e surface. Yet, any choice of a phase variable for
chaotic oscillators investigated in this paper leads to the same
macroscopic behavior \cite{josic01}.

With this phase definition, the phase dynamics can be described by
\begin{eqnarray}
A_{n+1} &=& {\mathcal M}(A_n),\\
d\phi /dt &=& \omega (A_n) \equiv \omega _{0} + \eta(A_n),
\label{phase-equ}
\end{eqnarray}
where the amplitude $A_n$ is the set of coordinates of the phase
point on the Poincar\'e surface at the nth intersection and
${\mathcal M}$ defines the Poincar\'e map that takes $A_n$ to
$A_{n+1}$. The ``instantaneous" frequency $\omega(A_n)= 2 \pi /
T_n$ is determined by the Poincar\'e return time $T_n =t_{n+1} -
t_n$ and depends in general on the amplitude. Assuming chaotic
behavior of the amplitudes, we can consider the term $\omega(A_n)$
to be the sum of the average (natural) frequency $\omega_0$ and
some effective ``noise" $\eta(A_n)$ with zero mean, although this
stochastic term has a purely deterministic origin \cite{pikovsky}.
Thus, Eq.(\ref{phase-equ}) has the solution
\begin{equation}
\label{phase-sol} \phi (t) = \phi_0 + \omega_0 t + \int^t_0 \eta
(\tau) d\tau.
\end{equation}
The variance of the integral $\int^t_0 \eta (\tau) d\tau$ is given
by
\begin{eqnarray}
    \left\langle \left( \int^t_0 \eta (\tau) d\tau \right)^2 \right\rangle &=&
    2\int^t_0 d\tau' (t-\tau') K(\tau') \nonumber \\
&\approx &2t\int^\infty_0 d\tau' K(\tau')\equiv 2t D_{\eta}.
\end{eqnarray}
Here, $K(\tau')= \langle \eta(\tau) \eta(\tau+\tau') \rangle$ and
the average $\langle \cdots \rangle$ is taken over the invariant
measure of the chaotic attractor. The approximate equality holds
provided that $\eta(\tau)$ has a finite correlation time $t_c$
such that $K(\tau') \approx 0$ for $\tau' > t_c$ and one considers
times $t \gg t_c$. This shows that the dynamics of the phase is
diffusive and the phase performs a random walk as long as $t_c$ is
finite and $D_{\eta}\neq 0$. However, as pointed out by Pikovsky
and co-workers \cite{pikovsky}, the dynamics of the phase
generally differs from stochastic Brownian motion because the
effective ``noise" cannot be considered as a Gaussian white noise
process. This was recently confirmed for a system of locally
coupled R\"{o}ssler oscillators where temporal correlations were
shown to exist \cite{davidsen02p}. These correlations determine
the speed of convergence of the time average $\bar{\omega
}(T)=T^{-1}(\phi (T)-\phi (0))$ toward $\omega _{0}$. The speed of
convergence
--- as measured by the standard deviation of the ensemble
distribution of $\bar{\omega}(T)$ --- scales as $1/T$. This is in
contrast to what one would expect if $\eta$ were a white noise
process where one obtains $1/\sqrt{T}$ scaling since
$\bar{\omega}(T) - \omega_0 = 1/T \int_0^T dt \eta (t)$ and the
standard deviation of $\int_0^T dt \eta (t)$ scales with
$\sqrt{T}$. Here, we provide further numerical and experimental
evidence that such a scaling generally occurs for oscillators
where the chaotic state is reached through a period-doubling
cascade. In particular, the type of coupling between the chaotic
oscillators seems to be irrelevant. Moreover, we show that the
temporal correlations induce an alternating behavior of
$K(\tau')$. This leads to an extremely small value of $D_{\eta}$
for these systems and suggests that they can generally be phase
synchronized.

\section{Globally Coupled R\"ossler Oscillators} \label{sec:sim}

\begin{figure}
 \vspace{1.3cm}
 \psfig{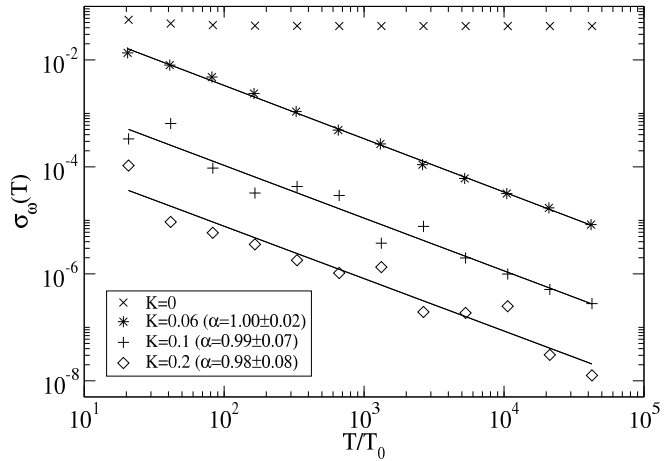}
 \caption{\label{scaling_l} Standard deviation
 $\sigma_{\bar{\omega}}(T)$ of the $\bar{\omega}(T)$ distribution
 in the phase synchronized state describing the speed of
 convergence toward $\omega_0$. The data points for $K=0.2$
 are shifted down by one decade for clarity.
 The Poincar\'e surface was chosen
 as the surface spanned by the negative $x_1$ axis and the $x_3$
 axis. The solid lines are best fits to the
 respective data and decay as $1/T^\alpha$.
 Note that $T_0 = 2 \pi /\omega_0$ depends slightly on $K$.}
 \end{figure}

We consider a $L \times L$ array of \emph{globally} coupled,
nonidentical chaotic R\"ossler oscillators:
\begin{equation}
\frac{\partial {\bf x}({\bf r},t)}{\partial t} = {\bf R}({\bf
x}({\bf r},t)) + K \sum_{{\bf\hat{r}} \in {\mathcal N}} \left ( {\bf
x}({\bf \hat{r}},t) - {\bf x}({\bf r},t) \right ),
\end{equation}
where $R_1 = -{\omega}({\bf r}) x_2 - x_3$, $R_2 = {\omega}({\bf
r}) x_1 + 0.2 x_2$, $R_3 = x_1 x_3 - 5.9 x_3 + 0.2$. The sites of
the lattice are labelled by ${\bf r}$, $K=K'/L^2$, $K'$ is the
coupling constant and ${\mathcal N}$ is the set of all sites on
the lattice. We take $L = 64$ and choose the ${\omega}({\bf r})$'s
randomly from a uniform distribution on the interval
$[0.99,1.01]$. This ensures that all oscillators are chaotic. We
find that the speed of convergence of $\bar\omega(T)$ toward
$\omega_0$ in the phase synchronized state scales as $1/T$ (see
Fig.~\ref{scaling_l}). This is also true for higher ${\omega}({\bf
r})$ dispersion.

\begin{figure}
 \includegraphics[width=\columnwidth]{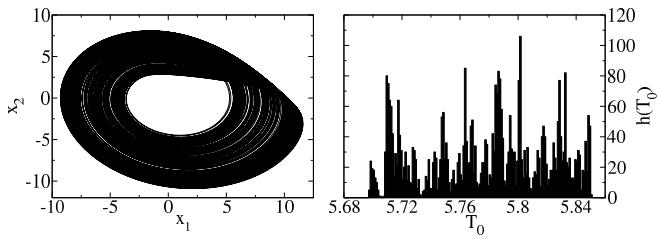}
 \caption{\label{attractor}Left panel: $x_1 - x_2$ projection
 of the local attractor at a single site of the lattice for
 $K=0.06$ in the phase synchronized state.
 All local attractors look similar to the one shown here. Right panel:
 A typical $T_0 = 2 \pi /\omega_0$ histogram $h(T_0)$ of
 the nonidentical chaotic oscillators for $K=0$, i.e., the distribution of
 the natural periods. The exact shape of the distribution
 depends on the $\omega({\bf r})$ realization. }
 \end{figure}

The $1/T$ scaling observed in the globally coupled R\"ossler
system follows from the fact that the local chaotic attractors,
although one-banded in the phase synchronized state (see
Fig.~\ref{attractor}), have internal structure.

\begin{figure}
 \includegraphics[width=\columnwidth]{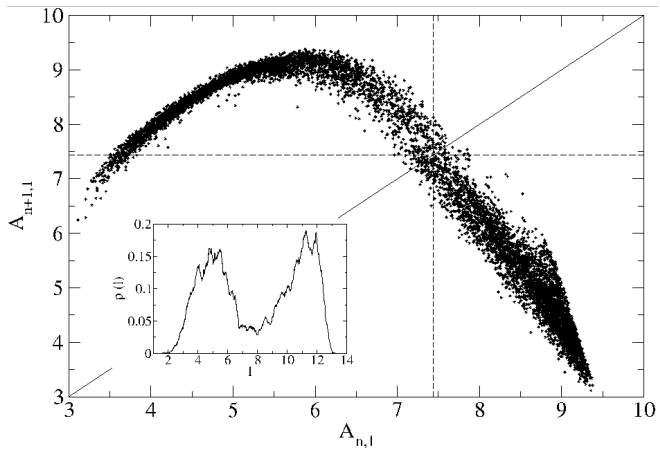}
 \caption{\label{nextamp_D006} Superposition of eight local
 next amplitude maps for $K=0.06$. The amplitude $A_{n,1}$ chosen
 here is the negative $x_1$ coordinate of $A_n$ because the $x_3$
 coordinate is almost constant and very close to zero. The dashed
 lines correspond to the intersection point of the bisectrix with
 the map. Inset: Density distribution $\rho (l)$ along the curve formed
 by the next amplitude maps.}
 \end{figure}

In the absence of coupling ($K=0$) the local non-identical
R\"ossler oscillators display a variety of banded chaotic
attractors because of period-doubling cascades and period-3
windows in their vicinity. In the phase synchronized state for
$K=0.06$, the local chaotic attractors are very similar to one
another and have an internal structure similar to that of a merged
two-banded attractor. Although this structure is not obvious in
the left panel of Fig.~\ref{attractor}, it can be seen clearly in
the local next amplitude maps (Fig.~\ref{nextamp_D006}) and in the
density distribution $\rho(l)$ (inset in Fig.~\ref{nextamp_D006}).
The intersection point of the map with the bisectrix defines four
regions in the map such that the two-bandedness of the attractor
can be characterized by the percentage $p$ of points in the upper
left and lower right quarter-planes. Indeed, only very few points
$(1-p=3\%)$ --- typically those for small $A_{n,1}$'s which are in
the lower left quarter-plane of the amplitude map --- stay in the
same region of the bimodal density distribution, implying that the
majority of the iterates map regions of high density on the left
to regions of high density on the right and vice versa.
This nearly two-banded character is also seen in the series of
$T_n$'s (see Fig.~\ref{instantT_D006}). An alternation of long and
short return times can be observed with rare exceptions. This
shows the close relation between topological characteristics (next
amplitude map) and temporal evolution ($T_n, T_{n+1}$ map)
and suggests the following explanation for the observed scaling:

 \begin{figure}
 \includegraphics[width=\columnwidth]{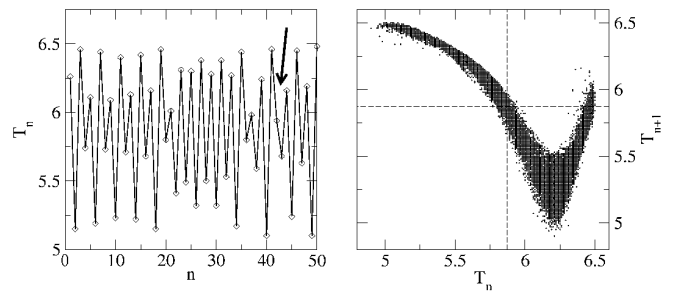}
 \caption{\label{instantT_D006}Time series of $T_n$ for $K=0.06$
 (left panel) and the corresponding $T_n, T_{n+1}$ map (right panel).
 The arrow highlights a deviation from the alternating behavior.
 These deviations occur with $1-p=2\%$ at this particular site
 of the lattice which is the percentage of points in the upper
 right and lower left quarter-plane in the lower panel.
 Note that only a short segment of the $T_n$ series is shown.}
 \end{figure}

Assuming that each internal "band" of the attractor is associated
with a distinct mean frequency $\omega_{1}$ or $\omega_2$ such
that $\omega_0 = 2 \omega_1 \omega_2 /(\omega_1 + \omega_2)$, and
the system alternates between bands, the "random" force in the
evolution equation (\ref{phase-equ}) for the phase of a single
oscillator takes the form
\begin{equation}
\eta(t)= (-1)^{\kappa+1} \frac{\omega_{\kappa}(\omega_1-\omega_2)}{(\omega_1+\omega_2)}
+\tilde{\eta}(t)\;,
\end{equation}
where $\kappa=1(2)$ when the system is on band 1(2). The noise
term $\tilde{\eta}(t)$ accounts for deviations of the period
within a band and deviations from the strict alternations between
bands. If $T$ is the time needed for $n$ oscillations, it follows
from Eq.(\ref{phase-sol}) that
\begin{eqnarray}
\bar{\omega}(T) - \omega_0 =
    \begin{cases}
        \int_0^T dt \frac{\tilde{\eta} (t)}{T}& \text{$n$ even},\\
        (-1)^{\kappa+1}\frac{2\pi (\omega_1-\omega_2)}{(\omega_1 +\omega_2)T} +
        \int_0^T dt \frac{\tilde{\eta} (t)}{T} & \text{$n$ odd},
    \end{cases}
\end{eqnarray}
where $\kappa=1(2)$ corresponds to band 1(2) appearing first in
the series. The term $\frac{2\pi (\omega_i-\omega_j)}{(\omega_1
+\omega_2)T}$ scales with $1/T$, while the term $\int_0^T dt
\frac{\tilde{\eta} (t)}{T}$ scales with $1/\sqrt{T}$ provided that
$\tilde{\eta}(t)$ is $\delta$-correlated noise. Since the
amplitude of the $1/\sqrt{T}$ term is very small, two scaling
regimes can be identified for the maximal deviation from
$\omega_0$: A $1/T$ scaling on intermediate time scales which is
dominated by the switching between the "bands" and a $1/\sqrt{T}$
scaling on long time scales. The intermediate time scale scaling
behavior can indeed be observed for each oscillator: The envelope
of $\bar{\omega}(T) - \omega_0$ decays as $1/T$. Note that the
standard deviation of the ensemble distribution of
$\bar{\omega}(T)$ decays as $1/\sqrt{T}$ for identical chaotic
R\"ossler oscillators and $K=0$ --- as expected from the law of
large numbers. In the case of nonidentical oscillators, phase
synchronization implies that the phases are synchronized such that
$\sqrt{\langle(\bar{\omega}(T) - \omega_0)^2\rangle}$, where the
average is taken over the ensemble, adopts the scaling of a single
oscillator reproducing the observed scaling and explaining the
results. This result does not depend on the type of coupling of
the oscillators as confirmed by the numerical results presented
here \cite{local} and in Ref. \cite{davidsen02p}. This explanation
is based exclusively on the fact that the series of the
instantaneous frequencies alternates with the exception of rare
events and that the system is phase synchronized.

\begin{figure}
 \includegraphics[width=\columnwidth]{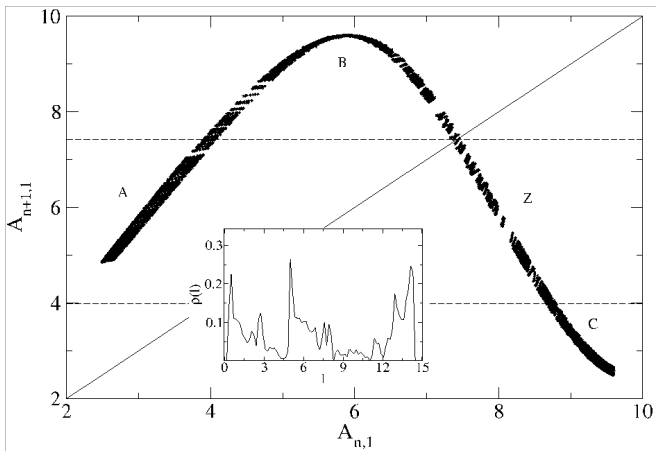}
 \caption{\label{nextamp_D02} Superposition of the next amplitude
 maps for $K=0.2$ at the same sites as in Fig.~\ref{nextamp_D006}.
 Inset: Normalized point density along the curve formed
 by the next amplitude maps. Three regions of high density can be
 identified.}
 \end{figure}

The above argument easily generalizes to attractors with multiple
internal ``bands" which we observe in the phase synchronized state
for higher coupling. Figure~\ref{nextamp_D02} shows that the
deviations from a three-banded structure are rather small for
$K=0.2$. Three bands A, B and C can be identified such that A is
mapped to B and C to A. However, band B is mapped to C and Z, and Z
back to B such that a pure three-banded structure does not exist.
The invariant measure of the map in area Z is very
low (approximately $10\%$).

The alternation of positive and negative values for $\eta(t)$
implies that $K(\tau')$ alternates as well. Hence, $D_{\eta}$ is
very small, i.e., a high degree of phase coherence persists. Our
findings suggest that this should be generally expected for
oscillators where the chaotic state is reached through a
period-doubling cascade. Thus, these oscillators would be phase
coherent suggesting that they can generically be phase
synchronized.

The change in the structure of the local attractors as the
coupling strength is changed is the analog of the
bifurcation structure seen in homogeneous locally coupled
R\"ossler oscillators where one finds period doubling different
from the isolated R\"ossler oscillator \cite{goryachev99}.

\section{Globally Coupled Electrochemical Oscillators} \label{sec:exp}
We now study the convergence of the time-averaged frequency in an
array of globally coupled electrochemical oscillators. A standard
three electrode electrochemical cell consisting of a nickel
working electrode array (64 1-mm diameter electrodes in an
$8\times 8$ geometry with 2 mm spacing), a
$\mathrm{Hg}/\mathrm{Hg}_{2}\mathrm{SO}_{4}/\mathrm{K}_{2}\mathrm{SO}_{4}$
reference electrode and a Pt mesh counter electrode were used. The
potentials of all of the electrodes in the array were held at the
same value ($V=1.310\, \mathrm{V}$) with a potentiostat (EG\&G
Princeton Applied Research, model 273). Experiments were carried
out in 4.5 M $\mathrm{H}_{2}\mathrm{SO}_{4}$ solution at a
temperature of $11\, ^{\mathrm{o}}\mathrm{C}$. The working
electrodes were embedded in epoxy and reaction takes place only at
the ends. The currents of the electrodes were measured
independently at a sampling rate of 100 Hz and thus the rate of
reaction as a function of position and time was obtained.

The electrodes were connected to the potentiostat through 64 individual
parallel resistors ($R_{\mathrm{p}}$) and through one series collective
resistor ($R_{\mathrm{s}}$). We employed a method of altering the
strength of global coupling while holding all other parameters constant.
The total external resistance was held constant while the fraction
dedicated to individual currents, as opposed to the total current,
was varied. A total resistance can be defined as
\begin{equation}
R_{\mathrm{tot}}=R_{\mathrm{s}}+R_{\mathrm{p}}/64\;.
\end{equation}
In these experiments $R_{\mathrm{tot}}=14.2\, \Omega $. The series
resistor couples the electrodes globally. The parameter
$\varepsilon $, the ratio of series to total resistance, is a
measure of the global coupling
\begin{equation}
\varepsilon =\frac{R_{\mathrm{s}}}{R_{\mathrm{tot}}}\;.
\end{equation}
For $\varepsilon =0$, the external resistance furnishes no
additional global coupling; for $\varepsilon =1$, maximal external
global coupling is achieved.

The population of chaotic oscillators is characterized by certain
amount of heterogeneity \cite{kiss02}. Without added coupling
there is a distribution of frequencies of the oscillators with a
mean of $1.219\, \mathrm{Hz}$ and a standard deviation of $18\,
\mathrm{mHz}$. With weak added global coupling ($\varepsilon
=0.1$) a nearly phase synchronized state occurs in which 63 of 64
oscillators have a frequency of $1.230\, \mathrm{Hz}$, and that of
the remaining element has a frequency of $1.237\, \mathrm{Hz}$.
The results are shown for this region of high phase synchrony.

\subsection*{Experimental Results}

\begin{figure}
\includegraphics[  width=0.95\columnwidth]{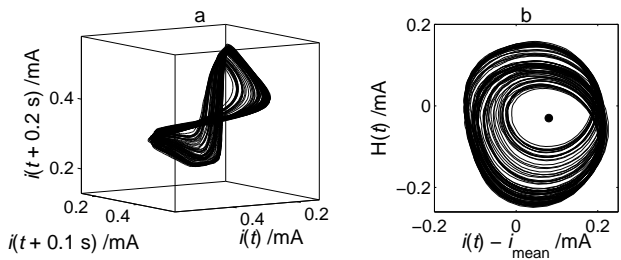}
\caption{\label{attr-exp}Experiment. Chaotic dynamics of a single
element at $\varepsilon =0.1$. a. Reconstructed attractor using
time delay coordinates. b. Two dimensional phase space
reconstruction using the Hilbert transform. The solid circle
represents the origin used for phase calculations. }
\end{figure}

The chaotic attractor of a representative single element of the
coupled system is shown in Fig. \ref{attr-exp}.
The chaotic state is adjacent to a period-three window and a
period doubling sequence. The reconstructed attractor (Fig.
\ref{attr-exp}a) is low-dimensional (correlation dimension $2.3\pm
0.1$). To obtain the phase, the Hilbert transform approach is
applied \cite{pikovsky}. The two-dimensional embedding using
Hilbert transform is shown in Fig. \ref{attr-exp}b. As in the case
of globally coupled R\"ossler oscillators, it is difficult to
distinguish the banded character of the attractor.

\begin{figure}
\includegraphics[  width=0.95\columnwidth]{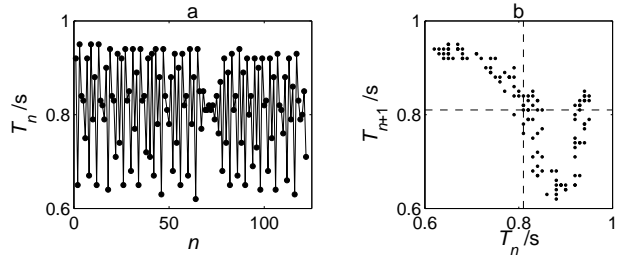}
\caption{\label{exp_tn} Experiment. Time series and maps of
discretized chaotic dynamics of a single element at $\varepsilon
=0.1$. a. Time series of return time ($T_n$) of the oscillations.
b. One dimensional map using the return time. The dashed lines
correspond to the average period of oscillations ($T_0=0.81\,
\mathrm{s}$) which in this case is identical with the intersection
of the map with the bisectrix.}
\end{figure}
The merged banded structure is more clearly seen in return maps
obtained from the series of return times ($T_n$). The series of
the return time clearly shows an alternation of long and short
oscillations (Fig. \ref{exp_tn}a). The map constructed using $T_n$
exhibits an approximately one-dimensional character (Fig.
\ref{exp_tn}b).

\begin{figure}
\includegraphics[  width=0.95\columnwidth]{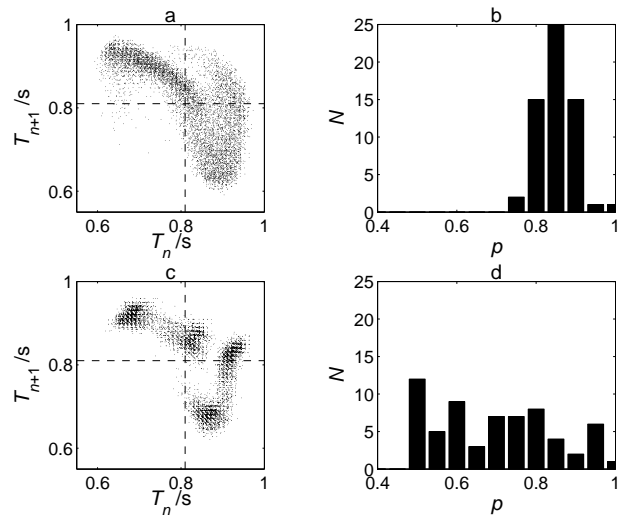}
\caption{\label{exp_twobanded} Experiment. Characterization of the
two-banded character of the population of sixty four oscillators.
Top row: $\varepsilon =0.1$. Bottom row: $\varepsilon =0$. a.
Superimposed one dimensional maps of the sixty-four sites. The
dashed lines correspond to the average period of oscillations
($\bar{T}=0.81\, \mathrm{s}$); $\varepsilon =0.1$. b. The
distribution of probability ($p$) of two-banded oscillations;
$\varepsilon =0.1$. c. Superimposed one dimensional maps of the
sixty-four sites; $\varepsilon =0.$ d. The distribution of
probability ($p$) of two-banded oscillations; $\varepsilon =0$.}
\end{figure}

The oscillators in the population are slightly different.
Figure~\ref{exp_twobanded}a
shows a superposition of the 1D maps of all the oscillators. %
Clearly, the majority of the phase points lie in the upper left
and lower right boxes indicating two-banded character. The
probability of two-banded oscillations varies from one site to the
other (Fig. \ref{exp_twobanded}b) with a mean of
$p_{\mathrm{m}ean}=0.85$ and a standard deviation of $0.05$. We
note that the two-banded character of the uncoupled system of
uncoupled oscillators ($\varepsilon =0$, $p_{\mathrm{m}ean}=0.7$,
Figs. \ref{exp_twobanded}c and d) is smaller than that of the
phase synchronized region. Therefore, during the transition to
phase synchronization with increasing the coupling strength the
two-bandedness observed in the time series of oscillations is more
pronounced.

As expected from the numerical simulations and theoretical
considerations, the strong (anti)correlations in the phase
dynamics have pronounced effects on the speed of convergence of
frequencies.
\begin{figure}
\includegraphics[  width=0.70\columnwidth]{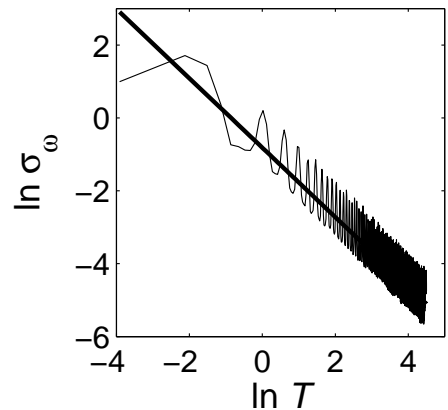}
\caption{\label{scaling-exp}Standard deviation of the
experimental frequency distribution as a function of the applied time average
(the thick line is a linear fit with slope -0.97) for $\varepsilon =0.1$.}
\end{figure}
Figure~\ref{scaling-exp} shows that the standard deviation of
frequencies is proportional to $1/T$ at $\varepsilon =0.1$.
(The fluctuations around the fitted line are caused by slowing down and
speeding up within a cycle.)  Similar scaling results are obtained for
other (larger) coupling strengths.

\section{Conclusion} \label{sec:conc}
For many systems of coupled oscillators, the chaotic state is
reached through a period-doubling cascade. This implies that a
Poincar\'e surface exists such that the next-amplitude map is
similar to that in Figs.~\ref{nextamp_D006} or \ref{nextamp_D02}.
As long as the correspondence between the next-amplitude map and
the relation of subsequent return times holds, this should lead to
a small value of $D_{\eta}$ and, in particular, to a $1/T$ scaling
in the phase synchronized state. The experimental results show
that such a structure of the series of return times can exists
even if no well-behaved next-amplitude map can be identified. Our
findings further suggest that chaotic oscillators where the
chaotic state is reached through a period-doubling cascade can
generally be phase synchronized due to the small value of
$D_{\eta}$, i.e., high degree of phase coherence. This is in
accord with findings in Ref.~\cite{pikovsky} where different
chaotic systems were analyzed with respect to their ability to be
phase synchronized. There it was found that the Lorenz system (where
the chaotic attractor is not reached through a period-doubling
cascade) cannot be phase synchronized. In particular, the next
amplitude map has a very different structure, i.e., a logarithmic
singularity.

Finally, in many biological and physical systems only short time
series are available.  Our results show that the determination of
frequencies in weakly coupled oscillatory systems may be
obtainable from such short time series.

\begin{acknowledgments}
This work (JLH and IZK) was supported in part by the National
Science Foundation (CTS-0000483) and the Office of Naval Research
(N00014-01-1-0603). RK and JD were supported in part by a grant
from the Natural Sciences and Engineering Research Council of
Canada. We thank Yumei Zhai for help with the experiments.
\end{acknowledgments}


\begin{thebibliography}{15}
\bibitem{local}
For local (next-neighbor) coupling, we obtain results
qualitatively similar to those shown in Figs.~\ref{attractor} and
\ref{nextamp_D006}, \ref{instantT_D006}.
\end{thebibliography}
\end{document}